\documentclass[prl,twocolumn,superscriptaddress,showpacs]{revtex4}
\usepackage{graphicx}
\usepackage{amsmath}
\usepackage{amsfonts}
\usepackage{bm}
\usepackage{color}

\newcommand{\Schroedinger}{Schr\"{o}dinger }
\newcommand{\ave}[1]{\langle #1 \rangle}

\begin{document}
\title{Relaxation dynamics of the Holstein polaron}

\author{Denis \surname{Gole\v z}}
\affiliation{J. Stefan Institute, 1000 Ljubljana, Slovenia}

\author{Janez \surname{Bon\v ca}}
\affiliation{J. Stefan Institute, 1000 Ljubljana, Slovenia}
\affiliation{Faculty of Mathematics and Physics, University of Ljubljana, 1000
Ljubljana, Slovenia}

\author{Lev \surname{Vidmar}}
\affiliation{J. Stefan Institute, 1000 Ljubljana, Slovenia}
\affiliation{Department of Physics and Arnold Sommerfeld Center for Theoretical Physics, Ludwig-Maximilians-Universit\"at M\"unchen, D-80333 M\"unchen, Germany}

\author{Stuart A. \surname{Trugman}}
\affiliation{Theoretical Division, Los Alamos National Laboratory, Los Alamos, New Mexico 87545, USA}

\begin{abstract}
Keeping the full quantum nature of the problem we compute the relaxation time of the Holstein polaron  after it was  driven far from the equilibrium by a strong  oscillatory  pulse.
Just after the pulse the polaron's kinetic energy increases and  subsequently exhibits  relaxation type decrease with simultaneous emission of phonons.
%We perform exact numerical calculation using full quantum coherence effects into account. 
In the weak coupling regime partial tunneling of the electron from the polaron self-potential is observed.
%in contrast to the intermediate coupling. No need to specially mention that. 
The inverse relaxation time is for small values of electron-phonon coupling $\lambda$ linear with  $\lambda$, while it deviates downwards from the linear regime at $\lambda \gtrsim 0.1/\omega_0$.
The imaginary part of the equilibrium self energy shows good agreement with the inverse relaxation time obtained from nonequilibrium simulations.
\end{abstract}

\maketitle

%\section{Introduction}
Photoexcitation  is due to  recent technological advances  in  ultrafast spectroscopy becoming one of the main experimental approaches to disentangle different elementary excitations in a real-time domain on a femtosecond scale.
 This rapidly developing field  enables a novel insight into quantum many-body systems, however it also requires the development of new theoretical concepts.
In this context, numerical simulations of nonequilibrium quantum mechanical systems may provide a key insight into processes on a femtosecond time scale.

Several pump-probe experiments reported the observation of self-trapping of excitions emerging after the pump pulse~\cite{dexheimer2000,sugita2001,morrissey10}. These experiments then
%In addition, two-photon photoemission experiments detected localization of electrons on a similar time scale~\cite{ge1998,miller2002}. 
triggered theoretical studies of different scenarios of the polaron formation, revealing a  complex interplay between a single electron and quantum phonons under nonequilibrium conditions~\cite{ku2007,fehske11}.

The role of electron-phonon (e-ph) interaction in several different classes of strongly correlated materials is despite intensive research still ambiguous since a subtle interplay between electron-electron and electron-phonon interaction may lead to various unconventional properties.
Pump-probe techniques have shown a potential to identify fingerprints of these interactions during the relaxation process~\cite{dalconte12,kim12,avigo12,kawakami10}.
One of the main challenges of such experiments represents the choice of underlying theoretical framework to interpret results.
Recently, the phenomenological two temperature model developed long time ago~\cite{kaganov1957,allen87}, was extended  to more involved approaches like the three-temperature model~\cite{perfetti07} and Boltzmann equation approaches \cite{demsar2003,kabanov2008,gadermaier10}.
%Another issue not clear at present is the role of temperature since in several phenomenological approaches relaxations times diverge with %decreasing temperature.
Moreover, the assumption of fast relaxation within the electron subsystem was recently challenged in the case of strong electronic correlations~\cite{eckstein11a}.
In contrast to many previous findings it was shown that in the $1D$ strongly correlated system coupled to phonons, the relaxation on short time scales is mostly due to e-ph interaction~\cite{matsueda11}.

Theoretical studies of the polaron motion in strong but constant  electric field started with the seminal work by Thornber and Feynman~\cite{thornber1970}. Later works mostly relied on the rate or Boltzmann equations~\cite{emin1987,rott2002,khan1987}. 
While  quantum coherent effects are absent in the Boltzmann description~\cite{wacker1999},  some recent approaches~\cite{bonca1997,vidmar_pol2011} show that taking fully into account quantum effects is decisive  to obtain proper electric field dependence of the drift velocity at large electric fields. Since most past works focused on the influence of a constant electric field on the Holstein polaron,  the impact of a short  oscillatory pulse on polaron relaxation dynamics remains an open problem despite significant advances in ultrafast spectroscopy \cite{bakulin2012,ge1998,morrissey10}.
%\cite{ge1998,miller2002,dexheimer2000,sugita2001}.

%\section{Model and numerical method}
In this Letter we present results of a fully quantum mechanical time evolution of the Holstein model driven far from the equilibrium  by a laser pulse. 
%A mayor challenge in investigation of electron-phonon coupled systems within the full quantum mechanical picture, is to  obtain  a complete  %solution of the relaxation process.
%We nevertheless show in this Letter that such solution is possible in the case of the Holstein polaron excited by a laser pulse.
%It enables us to follow the real-time energy transfer between the electron and phonon subsystem in two complementary time domains:
%{\it (i)} during the pulse, which is usually not incorporated into theoretical considerations based on a phenomenological level;
%{\it (ii)} for large times after the pulse when (almost) full relaxation within a closed system is achieved, i.e., in the time region where numerical simulations of many-body systems usually become inaccurate.
We determine characteristic relaxation times for the system which is initially in equilibrium at zero temperature. The investigated  Holstein polaron is subjected to a  spatially homogeneous and  time-dependent scalar potential that mimics a short laser pulse:
\begin{equation}
\phi(t)=A e^{-((t-t_{c})/t_{d})^2} \sin(\omega_{p} (t-t_{c})),
\label{phi}
\end{equation}
which is incorporated into the Hamiltonian via a Peierls substitution in the hopping amplitude:
\begin{eqnarray}
H &=& -t_0 \sum_{{l} ,\sigma} \left[ {\mathrm e}^{i \phi(t)}\; c^{\dagger}_{{l}, 
\sigma} c_{{l+1}, \sigma} + {\mathrm H.c.} \right] + {g} \sum_{{j}} n_{{j}} (a_{{j}}^\dagger + a_{{j}}) \nonumber \\
 & + & \omega_0\sum_{{j}} a_{{j}}^\dagger  a_{{j}},
  \label{ham}
\end{eqnarray}
where $c_{l,\sigma}^{\dagger}$ and $a_{i}^{\dagger}$ represent electron and phonon creation operators at site i, and $n_{i}=c_{i}^{\dagger}c_{i}$ is the electron density. $\omega_{0}$ denotes the 
dispersionless phonon frequency and $t_{0}$ is the nearest neighbor hopping amplitude.  

The system is described by two dimensionless parameters $\alpha$ and $\lambda$ where  $\alpha=\frac{\omega_{0}}{t_{0}}$ 
% determines which of the two subsystems (lattice or electron) is the fast and the slow one, while 
and $\lambda=g^2/(2\omega_{0}t_{0})$ that determine the crossover from adiabatic($\alpha \ll 1$) to nonadiabatic ($\alpha \gg 1$) limit and the weakly dressed electron ($\lambda \ll 1$) to a heavy polaron ($\lambda \gg 1$). 
We measure the electric field $F=-\partial_t\phi(t)$ in units of $t_{0}/e_{0} a$, where $e_{0}$ is the unit charge and $a$ is the lattice distance. The unit of energy is given by hopping amplitude $t_{0}$ and the unit of time is $\hbar/t_{0}$. From here on we set $a=e_{0}=\hbar=t_{0}=1$. Unless explicitly stated the phonon frequency is set to $\omega_{0}=1$. The pulse in Eq.~(\ref{phi}) is centered at  $t_c=5$, while the width is given by $t_d=2$.
%The energy difference between initial and temporary state is marked with $\Delta E(t)=\ave{H(t)}-\ave{H(t=0)}$ and similar for different parts of the Hamiltonian. 

We  solve the time-dependent \Schroedinger equation for a single Holstein polaron on an infinite one-dimensional chain.  We use  the numerical method based on the exact diagonalization of the variational Hilbert space  that led to numerically exact solutions of the polaron ground state~\cite{bonca1999} and low-lying excited-state properties~\cite{vidmar2010,barisic06}, as well as for description of the time-dependent case~\cite{vidmar_pol2011}.
%By systematically increasing the Hilbert space we were able to achieve high numerical precision of numerical results as well as sufficiently large propagation times that enabled accurate extraction of relaxation times of the model. 
The total energy gain from the external pulse is given by 
\begin{equation}
\Delta \ave{H(t)}=\int \ave{j(t)} F(t) dt,
\label{Eq:sumrule}
\end{equation}
where $j(t)=-\partial_\phi H$ is the current operator. Equation (\ref{Eq:sumrule}) is as well used as an additional  time  propagation accuracy check, see also Ref.\cite{golez2012}. 

%\section{Weak coupling relaxation}

\begin{figure}
\includegraphics[width=0.50\textwidth]{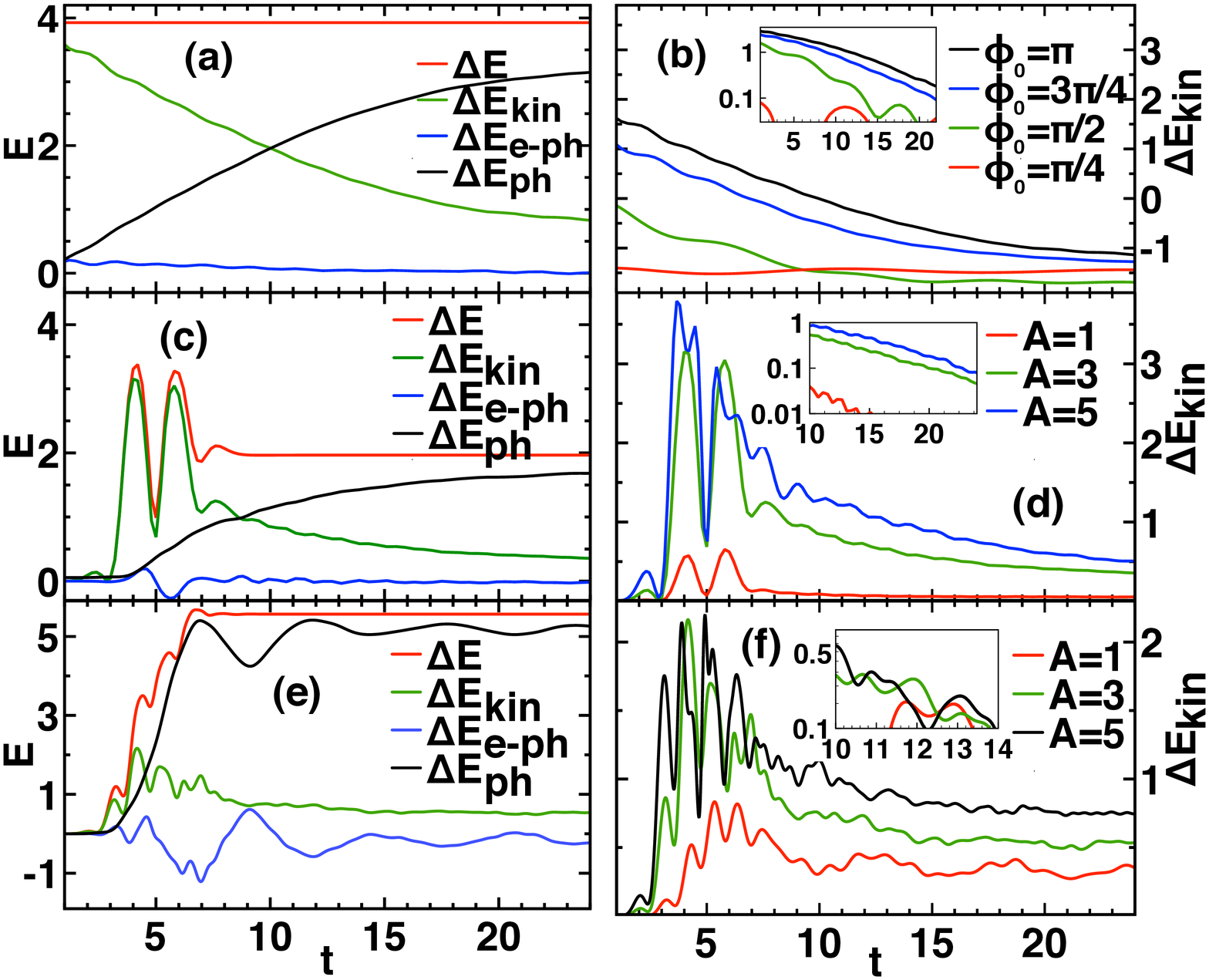}
\caption{(a) Expectation values of different parts of the Hamiltonian vs time at $\lambda=0.1$ and $\omega_0=1.0$, after action of  $\phi(t)=\Phi_0\theta(t)$, for $\Phi_0=\pi$. (b) $\Delta E_{\mathrm{kin}}(t)$ vs $t$  for different $\Phi_0=\pi,3\pi/4,\pi/2,\pi/4$, while other parameters are the same as in (a).
(c) Expectation values of different parts of the Hamiltonian vs time at $\lambda=0.1$ and $\omega_0=1$, after action of pulse as given in Eq.~(\ref{phi}) with $A=3$, where $\omega_{p}=1.5$ is set close to the maximum of the optical conductivity, while parameters $t_c=5$ and $t_p=2$ remain unchanged throughout this work.
(d)  $\Delta E_{\mathrm{kin}}(t)$ vs $t$  for different values of $A=1,3,5$, the rest is the same as in (c). The inset represent the difference between kinetic energy and $E_{kin}(t\rightarrow\infty)$ for the same pulse amplitudes as the main figure.
(e) Energy of the different parts of the Hamiltonian vs $t$ for  $\lambda=1.0$ and $\omega_{p}=2.5$.
(f) $\Delta E_{\mathrm{kin}}(t)$ vs $t$  for different values of $A=1,3,5.$ while the rest is the same as in (e). 
%The inset represent the difference between kinetic energy and $E_{kin}(t\rightarrow\infty)$. 
 }
\label{Fig:energy_weak}
\end{figure}

Before examining the more physically relevant case we analyze the action  of a pulse in a form of a delta function $F(t)=-\Phi_0 \delta (t)$ that  leads to a simple form of the scalar  potential $\phi(t)=\Phi_0 \theta(t)$ where $\theta(t)$ is the Heaviside function. Action with  $\phi(t)$ on a free electron  state ($\lambda=0$) at $k=0$   shifts its kinetic energy from $E_{\mathrm{kin}}=-2$ at $t<0$ to $E_{\mathrm{kin}}=-2\cos(\Phi_0)$ when  $t>0$, leaving the electron in an excited, but an eigenstate at $t>0$.  In Fig.~\ref{Fig:energy_weak}(a) we show results for $\lambda=0.1$ and $\Phi_0=\pi$. Just after the pulse, the increase of the total  $\Delta E_{\mathrm{}}(t)=\ave{H(t)}-\ave{H(t=0)}$, as well as the kinetic energy $\Delta E_{\mathrm{kin}}(t)$, reach  value  $\Delta E_{\mathrm{}}\sim \Delta E_{\mathrm{kin}}\sim -2\cos(\Phi_0)+2=4$ while the change of the phonon energy $\Delta E_{\mathrm{ph}}$ remains close to its value in the polaron ground state, {\it i.e.} $\Delta E_{\mathrm{ph}}\sim 0$.
After initial time $t\gtrsim t_i\sim 10$, $\Delta E_{\mathrm{kin}}(t)$ exhibits  a relaxation type  exponential decay towards a constant value $\Delta E_{\mathrm{kin}}(t\to\infty)$, meanwhile $\Delta E_{\mathrm{ph}}(t)$ increases and the electron-phonon interaction term $\Delta E_{\mathrm{e-ph}}(t)$ remains nearly a constant. This dynamics is interpreted as a transfer of the excited electron kinetic energy into phonon excitations.
%Results are consistent with a transfer of the excited electron kinetic energy into phonon excitations.
 In Fig.~\ref{Fig:energy_weak}(b) only  $\Delta  E_{\mathrm{kin}}(t)$ is shown for different choices of pulse amplitude $\Phi_0$. In all cases except for  $\Phi_0=\pi/4$, $\Delta E_{\mathrm{kin}}(t)$ decrease exponentially with roughly the same relaxation time. In the case of  $\Phi_0=\pi/4$ the increase of $\Delta E_{\mathrm{kin}}(t)$ is lower than $\omega_0$ indicating that the energy transfer from the pulse was insufficient to allow electron  relaxation via phonon emission, thus no relaxation is observed.

%\begin{figure}
%\ includegraphics[width=0.22\textwidth]{figures/fig5a.eps}
%\includegraphics[width=0.22\textwidth]{figures/fig5b.eps} \\
%\includegraphics[width=0.22\textwidth]{figures/fig5c.eps} 
%\includegraphics[width=0.22\textwidth]{figures/fig5d.eps}
%\caption{ The inset represent the difference between kinetic energy and $E_{kin}(t\rightarrow\infty)$. (c) Energy of the different parts of the Hamiltonian as a function of time in the intermediate coupling regime, namely $\lambda=1.0$ (d) The difference in the kinetic energy as a function of time for different amplitudes of the pulse, $A=\pi,3\pi/4,\pi/2,\pi/4$.}
%\label{Fig:energy_quench}
%\end{figure}

We next consider a more realistic form of a pulse described by the scalar potential of  Eq.~(\ref{phi}).  
At small $\lambda=0.1$  we observe a gain of the  total energy $\Delta E_{\mathrm{}}(t)$, see Fig.~\ref{Fig:energy_weak}(c), signaling that despite rather weak $\lambda$, the system has absorbed a substantial amount of energy. Note that after the pulse is switched off for $t\gtrsim 9=t_{\mathrm{off}}$, the total energy remains constant while there is a clear redistribution between expectation values of the parts of the Hamiltonian. Redistribution between the kinetic $\Delta E_{\mathrm{kin}}(t)$ and the phonon $\Delta E_{\mathrm{ph}}(t)$ part of the total energy  clearly indicates a relaxation of the system. %Different parts add up to a constant for $t>t_{\mathrm{off}}$:  $\Delta E_{\mathrm{}}(t) =  \Delta E_{\mathrm{kin}}(t) +  \Delta E_{\mathrm{e-ph}}(t) +  \Delta E_{\mathrm{ph}}(t) $, see also Eq.~(\ref{ham}). 
After the pulse,  $\Delta E_{\mathrm{ph}}(t)$ increases with time and  $\Delta E_{\mathrm{kin}}(t)$ exponentially decreases, while $\Delta E_{\mathrm{e-ph}}(t)$ oscillates  roughly around zero. This behavior is similar to the case when $\phi(t)=\Phi_0\theta(t)$, see Fig.~\ref{Fig:energy_weak}(a).  For $t\gg t_{\mathrm{off}}$ most of the gained energy is absorbed by the lattice. We should remark that the relaxed expectation values of the kinetic energy are not the same as before the pulse. As a general rule we find in all other cases $\Delta E_{\mathrm{kin}}(t\to\infty)\lesssim \omega_0$, which indicates that a finite value of $\Delta E_{\mathrm{kin}}(t\to\infty)$  is a consequence of the gap for optical phonons~\cite{wacker1999}. Nevertheless, we detect a clear tendency of the exponential decay of the kinetic energy towards only a slightly elevated value $\Delta E_{\mathrm{kin}}(t\to\infty)$ in comparison to the initial energy $E_{\mathrm{kin}}(t=0)$. 
%Predominant absorption of the excess energy in to lattice  is expected since only lattice degrees of freedom represent in principle an infinite %reservoir for the absorbed energy. 
We have also computed  the relaxation of the kinetic energy for different amplitudes of the pulse,  presented in Fig. \ref{Fig:energy_weak}(d). We found that the decay time is within our numerical accuracy independent of the amplitude of the pulse $A$ as long as $A$ exceeds a threshold value $A\gtrsim 1$, see also the inset of Fig.~\ref{Fig:energy_weak}(d), which is in agreement with the Boltzmann theory. 

%Here  we should note that the system dimensionality plays crucial role in static properties of the polaronic systems \cite{emin1976,ku2002}. 
Our results are  qualitatively consistent with experiments on quasi-one-dimensional systems~\cite{rashba1982}, where exponential relaxation takes place \cite{dexheimer2000,sugita2001}. Relaxation was also observed in $2D$ systems~\cite{ge1998,ge2000,miller2002}, where it is a consequence of the molecular dipoles rearrangement in the vicinity of the electron, which can be described by formally equivalent electronic polaron model \cite{mahan2000}.

%Using analogy with the analysis in Ref.\cite{ku2007}, the probability amplitude to relay into the lowest band ( mark that due to dispersion-less phonons this need not be $k=0$ point of the equilibrium model ) is to the first order in $\lambda$ proportional to the $|\braket{\phi(t)}{c^{\dagger}_{k}0}|^{2}$, which is just absolute value of the Fourier transform of the spectral function. In the weak coupling limit the spectral function is composed of the isolated delta peak at $\omega=0$ with weight $a$, which corresponds to the ground state and Lorentzian positioned at $\omega_{1}>\omega_{0}$ with weight $b$, whose width is corresponds to the inverse relaxation time $1/\tau$, see Fig. 2 in Ref.\cite{ku2007}. The probability for relaxation into the ground state is therefore:
%\begin{equation}
%P(t)=a^{2}+b^{2} \exp^{-2 t/\tau}+2 a b \exp^{-t/\tau} \sin(\omega_{1} t).
%\end{equation}
%This form shows that the slowest exponential decay is multiplied by a sin oscillation with the frequency corresponding to the energy difference between the ground state and the center of the Lorentzian in the spectral function.  This is actually in good agreement with the relaxation of the kinetic energy ( lowering of the kinetic energy comes from the scattering on the phonon) composed of the  exponential envelope and oscillations with frequency $\omega_{1}>\omega_{0}$, see Fig.\ref{Fig:energy_weak}(b). (jaz tu ne bi kompliciral s tem kaksna je renormalizacija effektivne frekvence\ldots)

%\section{Intermediate coupling regime}
In the intermediate coupling regime, namely $\lambda=1.0$,  a different response of the system is expected due to a  bigger gap between the polaron band and the continuum of excited states. 
%The response of the system strongly depends on the frequency of the incoming photon $\omega_{p}$. 
After the pulse almost the entire excess  energy is absorbed into the lattice vibrations, {\it i.e.}  $\Delta E_{\mathrm{}}(t>t_{\mathrm{off}})\sim \Delta E_{\mathrm{ph}}(t>t_{\mathrm{off}})$, see Fig.~\ref{Fig:energy_weak}(e). In addition to  a large increase, the latter displays as well  oscillations with the period corresponding roughly to the phonon frequency. 
%Predominant absorption of the excess energy in to lattice  is expected since only lattice degrees of freedom represent in principle an infinite reservoir for the absorbed energy. 
%Nevertheless, we still observe increase in the kinetic energy and its exponential relaxation after the pulse.
%This strong oscillation are purely coherence effects and we need to take into account full quantum mechanical description of the system. Despite the oscillation the relaxation (in average) is exponential and we can once more define the relaxation time, which is still independent on the pulse amplitude, although were are far beyond Boltzmann ( or rate-equation) description. 
 In Fig.~\ref{Fig:energy_weak}(f) we present results of  $\Delta E_{\mathrm{kin}}(t)$ for different amplitudes of the pulse. Apart from  pronounced oscillations, we again observe approximately exponential decay in the kinetic energy, roughly independent of the strength of $A$ when $A\gtrsim 1$. We also  observe distinct long-time limits of kinetic energies, however in all cases $\Delta E_{\mathrm{kin}}(t\gg t_{\mathrm{off}}) < \omega_0$.
%{\color{green}  Ask Trugman for the analysis of the strong coupling transition rates.}
 
{\color{red}
%When the  electron-phonon coupling is increased even further, for   $\lambda \ge 1$ (not shown),  the system exhibits only slightly damped %oscillatory dynamics on a very long time scale. Defining relaxation time for $\lambda \ge 1$ becomes increasingly more ambiguous.  The main %reason is that the relaxation time becomes comparable to the duration of the pulse, {\it i.e.} $\tau\sim 2t_d$. 
}
\begin{figure}[h]
\includegraphics[width=0.43\textwidth]{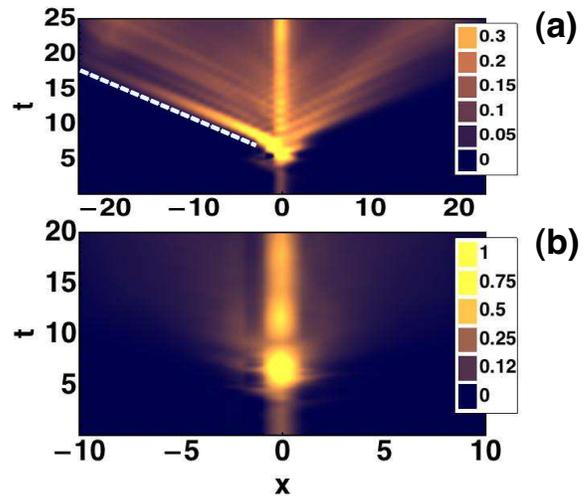}
\caption{Electron-phonon correlation function $\gamma(x)$ for pulse amplitude $A=3$; (a)  $\lambda=0.1$ and the pulse frequency $\omega_{p}=1.5$ and (b)  $\lambda=1.0$ and  $\omega_p=2.5$,  at different times. Note that there are different scales on both figures.}
\label{Fig:corr}
\end{figure}

To get further insight into the relaxation dynamics we calculated the average number of phonon quanta located at a given distance $x$ from the electron
\begin{equation}
\gamma(x)=\langle \sum_{i}n_{i}a_{i+x}^{\dagger}a_{i+x} \rangle,
\end{equation}
satisfying  the sum rule $\ave{n_{ph}}=\sum_{x}\gamma(x)$. At time $t=0$, $\gamma(x)$ displays a pronounced peak at the position of the electron $(x=0)$, consistent with the shape of the polaron in equilibrium. 
In the weak coupling regime, at $\lambda=0.1$ and $t>t_{\mathrm{off}}$, $\gamma(x)$ shows beside original polaron correlation peak also pronounced peaks  separating from the central peak in both directions as time increases, see Fig.~\ref{Fig:corr}(a).  This result is consistent with a hypothesis that a strong pulse splits the polaron into an excited polaron and a nearly free electron. The wavefunction is a superposition of an excited polaron, responsible for the large value of $\gamma(x)$ at $x=0$  and a nearly free electron, traveling predominantly in the $x>0$ direction. The more pronounced signal for $\gamma(x<0)$ can be interpreted as a partial tunneling of electron part of the wavefunction from polaron self-potential that remains located at $x=0$. The asymmetry on the parity transformation $x\rightarrow-x$ is dynamically induced and can be tuned by choosing different shape of the incoming pulse.

%, thus escaping its potential well that remains located at $x=0$, and  giving rise to a strong signal in $\gamma(x)$ at $x<0$. A weaker signal of $\gamma(x)$ for $x>0$ can be interpreted as coming from the part of the nearly free electron wavefunction, traveling in $x<0$ direction.   The more pronounced signal for $x<0$ can be interpreted as a partial tunneling of electron part of the wavefunction from the polaron self-potential. 
%The asymmetry of the correlation function $\gamma(x)$ on $x=0$ for initial times corresponds %to the fact that partially tunnelled part of the electron wavefunction is a bare electron, which %starts to form new "partial polaron",
%whose formation is seen as first pronounced peak for $x>0$. 
The peak at $x<0$  starts to diminish with time because: {\it (i)}  the escaped nearly free electron is gradually captured by the lattice and {\it (ii)} the excited electron redistributes its excess energy into constantly spreading  area of excited lattice vibrations giving rise to nearly uniform but elevated values of $\gamma(x)$, clearly seen in Fig.~\ref{Fig:corr}(a). 
%
%due to: (a) continous partial tunnelling of the electron from the original polaron self potential, %(b) additional electron phonon scattering occur
%which distort the lattice. 
%
Here we should stress that redistribution in the correlation function must be due to electron's motion since phonons are dispersionless. This hypothesis is well supported by the estimation of the velocity of the side peaks $v_p=\Delta x/\Delta t\sim 2$ representing the maximal group velocity of the  weakly  coupled  electron [see dashed line in Fig.~\ref{Fig:corr}(a)].  Similar  partial tunneling was noticed within the adiabatic limit of the driven Su-Schrieffer-Heeger problem \cite{johansson2004}. Above considerations give complementary real space interpretation of related experimental results, where excitations are identified in the frequency domain~\cite{dexheimer2000,sugita2001}.
%and diminishing of the side peak intensity
%represents the electron phonon scattering. 
%The more pronounced signal  for $x<0$  can be interpreted as  the polaron "self-potential", moving away from the polaron peak and diminishing in its intensity. In reality, since phonons are dispersionless, the "self potential" does not move anywhere, it is the polaron that distances itself from the "self potential". This hypothesis is well supported by the estimation of the velocity $v_p=\Delta x/\Delta t\sim 2$ representing the maximal group velocity of the  weakly  coupled  electron.  This partial tunneling was noticed within the driven Su-Schrieffer-Heeger problem \cite{johansson2004}, where the phonon degrees of freedom are treated classically and corresponds to the adiabatic limit $\omega_{0}/t \ll 1$. 
%In Fig.~\ref{Fig:corr}(a) we also observe that in addition to the "self potential" there are other phonon excitations, generated by the optical pulse, that spread  away from the polaron position.  Such redistribution of phonon excitations  must be due to polaron motion since phonons are dispersionless.  
% the $\gamma(r)$ after the pulse $t>t_{\mathrm{off}}$ shows two pronounced peaks, where 
%the peak at the origin corresponds to extra phonon excitations constituting  the excited polaron state. The second peak can be interpreted as
%a leftover phonon excitations  due to  partial tunneling of the electron from the polaron "self-potential",   moving away from the polaron peak and diminishing in its intensity.  
%
In the strong coupling regime, namely $\lambda=1.0$ as presented in Fig.~\ref{Fig:corr}(b), the polaron peak at $\gamma(x\sim 0)$ is preserved, but it broadens with time and we observe no peak due to partial electron tunneling. 
%(Poglej, ce morda pri vecjih frekvencah pulza ne bi zbezal iz jame... Nekako mu je potrebno dodati dovolj energije za tuneliranje). 

\begin{figure}
\includegraphics[width=0.45\textwidth]{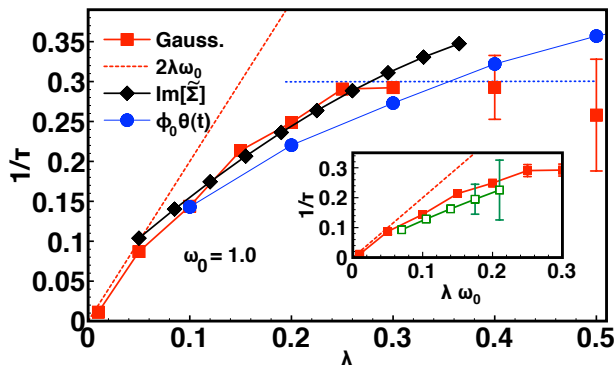}
\caption{Inverse relaxation time $1/\tau$ (squares) vs $\lambda$ for $\omega_{0}=1.0$ using Gaussian pulse,  Eq.~(\ref{phi}). The dashed line represents  $1/\tau = 2\omega_0\lambda$, see discussion in text. Diamonds: the imaginary part of the averaged equilibrium self energy ${\rm Im}[\tilde \Sigma]$. Circles: $1/\tau$  after the instantaneous pulse $\phi(t)=\Phi_0 \theta(t)$ (circles) where pulse strength $\Phi_0=\pi$ was used. The horizontal dotted  line indicates  the inverse pulse width. The inset presents $1/\tau$ vs. $\lambda\omega_0$, using Gaussian pulse,  Eq.~(\ref{phi}), for $\omega_0=1$ (full squares) and $\omega_0=0.7$ (open squares).}
\label{Fig:relaxation}
\end{figure}

We computed the relaxation time by fitting the expectation value of the kinetic energy after the pulse with the simple expression $\Delta E_{\mathrm{kin}}(t)=   \Delta E_{\mathrm{kin}}(t\to \infty) +B e^{-t/\tau}$, where $ \Delta E_{\mathrm{kin}}(t\to \infty)$ is the kinetic energy after relaxation  and $\tau$ is the relaxation time, see Fig.~\ref{Fig:relaxation}. The inverse relaxation time $1/\tau$ in the extreme weak coupling $\lambda \le 0.1$ regime shows linear increase with electron-phonon coupling, consistent with   $1/\tau \approx 2 \omega_{0} \lambda$. Considering an emission of a phonon by the excited electron with the kinetic energy above one-phonon threshold using Fermi golden rule yields linear dependence on $\lambda$,  {\it i.e.} $1/\tau = 2 \omega_0\lambda/\sin(k_f)$, where $k_f$ is the final electron's momentum and the inverse relaxation time is determined by the longest decay time, namely at $k_{f}=\pi/2$. 
With increasing $\lambda$ the inverse relaxation time $1/\tau$ first deviates downwards from the linear $\lambda$ dependence and  then  saturates as it becomes  comparable to the pulse width. The larger error bars are a consequence of  strong oscillations and a smaller spatial extend of the variational phase space in the strong coupling regime. Calculating $1/\tau$ at smaller $\omega_0=0.7$ shows that the scaling of $1/\tau\sim f(\lambda\omega_0)$, where $f(x)$ is some unknown function [note that within Fermi golden rule $f(x)=2x$],  persists beyond the linear in $\lambda$ regime, see the inset of Fig.~\ref{Fig:relaxation}. The saturation of $1/\tau$ is absent  in the case when  $\phi(t)=\Phi_0\theta(t)$, also presented in Fig.~\ref{Fig:relaxation}.  The relaxation time after an instantaneous pulse corresponds to the process averaged over all frequencies and this explains the deviation of the inverse relaxation time from a finite width pulse.
%The general dependence of the relaxation time for smaller phonon
%frequencies, namely $\omega_{0}=0.7$, shows similar behaviour with the electron-phonon coupling $\lambda$, as shown in the inset of Fig. \ref%{Fig:relaxation}.

An alternative method for the computation of $1/\tau$ from equilibrium properties is via the imaginary part of the self energy ${\rm Im}[\Sigma(\omega)]$ \cite{sensarma2010} that represents the inverse of the relaxation time of the quasiparticle excitation $1/\tau_\Sigma={\rm Im}[\tilde \Sigma(\omega_{p})]$. 
%The self energy was calculated using continued fraction expansion and Kramers-Kronig relations on the ground state in the equilibrium \cite{dagotto1994}. 
Since  ${\rm Im}[\Sigma(\omega)]$ depends on the value of the frequency of the pulse $\omega_p$ we define $1/\tau_\Sigma$ via the  average value  $\tilde \Sigma(\omega_{p})=(1/\sqrt{2 \pi \sigma}) \int  e^{-(\omega-\omega_{p})^2/(2 \sigma)^2} \Sigma(\omega)\mathrm{d}\omega,$ where $\sigma=1/t_{d}$, and integrate  over a distribution  of frequencies corresponding to  the Fourier transform of the pulse.  Although there is no a priori reason that this equilibrium quantity represents the inverse relaxation time even after the strong non-equilibrium process the agreement with the actual values of the inverse time is excellent in the weak coupling limit.  %The deviation of the inverse relaxation time $1/\tau$ extracted from simulation and imaginary part of the self energy for $\lambda > \lambda_{c}$ is consequence of the stronger inter quasi-particles excitation interaction.
%It is instructive to note that $1/\tau_\Sigma$  correctly captures the downwards deviations from the linear dependence on $\lambda$.
%Our simulations show that  the inverse relaxation time is rather independent on the frequency $\omega_p$ of the pulse, while  the imaginary %part of the equilibrium self energy shows rather pronounced peaks at $\omega=\omega_{0},2\omega_{0}$. This may be due to a finite duration %of the pulse and its Gaussian envelope that leads to a distribution of different frequencies in its Fourier transform. 

%%%DENIS:This is slightly dangerous to mention. Referee might get an idea to ask us to compute the out-of-equilibrium Green's function... 
%Another important aspect  is that during the non-equilibrium process the self energy can change in the nontrivial way, however since we were  not able to calculate the non-equilibrium self energy of the system this issue remains an open question. 
 % (Preveri se enkrat kako je z relaxacijo v SC limiti pri manjsi frekvenci pulza, kjub temu, da ni velikega odziva Sem pregledal in vse zgleda precej podobno. Problem je pri $\lambda=0.1$ kjer nakloni res kazejo nekaj odklona, vendar je lahko to posledica prekratkih casov propagacije in potem ni pravilno dolocena dolgocasovna kineticna. sem dal racunat.	 ).

%\begin{figure}
%\includegraphics[width=0.45\textwidth]{figures/fig4a.eps}
%\includegraphics[width=0.45\textwidth]{figures/fig4b.eps}
%\caption{ Absorbed energy into the system divided by the square amplitude  of the pulse $A^{2}$, with lower frame representing added change of energy in linear response theory. (b) Absorbed energy into the system divided by the square amplitude  for certain frequency of the pulse.}
%\label{Fig:energy}
%\end{figure}

%\section{Conclusion}
%
To conclude, in this Letter we studied the relaxation dynamics of the Holstein polaron after the strong photo-excitation. In all cases a threshold value of the absorbed energy $\Delta E \sim \omega_0$ exists above which the relaxation dynamics via phonon emission  is observed. 
We computed the relaxation time that is  mostly  independent of the shape of the pulse.
We focused on the values of e-ph coupling $\lambda$ below the crossover to the strong-coupling regime, {\it i.e.}  $\lambda_c \sim 1$.
%Note, \lambda_c is in general independent of \omega_0!!!
In this range of $\lambda$, relaxation dynamics exhibits two distinct regimes with qualitatively different behavior:
the regime of very weak e-ph coupling, $\lambda  <0.1/\omega_0$, and
the regime when $ \lambda\gtrsim 0.1/\omega_0$.

In the weak-coupling regime $\lambda < 0.1/\omega_0$,  $1/\tau$  roughly follows the linear scaling $1/\tau \sim 2\omega_0\lambda$ obtained from the Fermi golden rule. The relaxation process is described by the lowering of the kinetic energy at the expense of the increased phonon energy.
%, {\color{blue} while no significant oscillations in $\Delta E_{\mathrm{kin}}(t)$ are observed. 
A  partial tunneling of the electron from the polaron self-potential occurs, resulting in a rapid increase in time of the average distance between the electron and lattice deformation, followed by the subsequent re-trapping process. 
Moreover, values of $\tau$ obtained from real-time calculations can be as well reproduced by calculation of the imaginary part of the averaged electron self-energy. 
%The relaxation of the Holstein polaron for $\lambda \ll 1$ is characterized by a relatively large extend of lattice deformation as well as large relaxation times.
%Such behavior is observed up to $\lambda < \lambda^* \sim 0.3$.
%We should note that to  account for the full relaxation process in this regime by means of numerical methods, it is necessary to construct a large functional space where the maximal distance between electron and phonon should strongly exceed the size of the polaron in equilibrium.

In the regime of $ \lambda \gtrsim 0.1/\omega_0$,  $1/\tau$ deviates from the linear in $\lambda$ dependence  well below the crossover to small polaron regime. 
Most of the energy absorbed by the system is again deposited into phonon excitations, however, in contrast to $\lambda\ll 1$, phonon excitations remain in the close proximity of the polaron.  
The real-time calculation reveals oscillations in $\Delta E_{\mathrm{kin}}(t)$ and other expectation values, with the period of the phonon frequency, $T \sim 2\pi/\omega_0$.
This result is in agreement with a recent study of a half-filled $2D$ Hubbard-Holstein model~\cite{defilippis12}, where oscillations with the period $T$ were observed for $\lambda < \lambda_c$.
%Even though the latter study was performed on a rather small cluster of 8 sites, comparison with our data indicated that these oscillations are %not finite size effect.

%In our study we have showed that the Holstein polaron system is at this moment the most complex quantum e-ph coupled system where the %numerically accurate full relaxation process, in particular the determination of relaxation times in the long-time limit, $t/t_d \gg 1$, can be %achieved. 

In comparison with the experiments our relaxation times are short, which is a consequence of rather high adiabatic coefficient $\alpha=1$. 
While in the regime $\lambda\lesssim 0.1/\omega_0$ the effect on $\tau$ by lowering $\alpha$ can be obtained from the Fermi golden rule  $\tau \propto\ 1/\lambda\omega_0=1/\lambda\alpha t_0$, our numerical results show that even in the regime when $\lambda\gtrsim 0.1/\omega_0$  there exists approximate scaling $\tau\sim 1/f(\lambda\omega_0)$ that can be used to extend our results towards potentially more physically relevant values of $\alpha$ and consequently longer relaxation times $\tau$.  Another interesting topic for further research is the effect of the dimensionality, since our simulations agree with the exponential relaxation in quasi $1D$ system, while relaxation is slower in $2D$ experiments.

Let us briefly discuss the possible relevance of our results to correlated electron systems with a finite electron density, such as {\it e.g.} the Hubbard model.
Recent nonequilibrium DMFT study showed that the relaxation time of the pump--excited Hubbard model in the case of large Coulomb repulsion is unexepctedly long~\cite{eckstein11a} due to exponentially slow decay rate of pump-generated doublons, in agreement with experiments on optical lattices~\cite{strohmaier10}.
These results open  a relevant question about the dominant mechanism of fast relaxation observed in photoexcited strongly correlated materials.
Lately, the emission of phonons in the $1D$ Hubbard-Holstein model was indeed shown to be a very efficient relaxation mechanism~\cite{matsueda11}, where a non negligible amount of phonons is already emitted during the application of the pulse.
This observation is in agreement with our results presented in Fig.~\ref{Fig:energy_weak}, and suggests that the time-evolution within the full nonequilibrium process [both during and after the pulse] may provide a comprehensive understanding of the photoexcited polaronic systems.
%However, simulations of the full relaxation of electron-phonon coupled systems within the full quantum mechanical picture is numerically hard to achieve when dealing with finite electron densities.

%\acknowledgements
\begin{acknowledgments}
Stimulating discussions with T. Tohyama, V. V. Kabanov and C. Gadermaier are acknowledged.
J.B. and L.V. acknowledge support by the P1-0044 of ARRS, Slovenia.  J.B expresses gratitude for the  support of   CINT user program, Los Alamos National Laboratory.
\end{acknowledgments}

\bibliography{Polarons.bib,Books.bib}
\end{document}